\documentclass[final,5p,times]{elsarticle}

\usepackage{graphicx}
\usepackage{color}
\usepackage{amsthm}
\usepackage{lineno}
\usepackage[abs]{overpic}
\usepackage{amssymb}
\usepackage[tbtags,sumlimits]{amsmath}

\journal{EPJ C}

\begin{document}



\title{Flux measurement of fast neutrons in the Gran Sasso underground laboratory}

\author{Gianmarco Bruno\corref{cor1}\fnref{label1,label4}}
\cortext[cor1]{\textit{Email address}: gianmarco.bruno@nyu.edu}

\author[label1,label3]{Walter Fulgione}


\address[label1]{INFN-Laboratori Nazionali del Gran Sasso, Via G. Acitelli 22, I-67100 Assergi (L'Aquila), Italy}
\address[label4]{\emph{Present Address:} New York University Abu Dhabi, PO Box 129188, Abu Dhabi, United Arab Emirates}
\address[label3]{INAF, Osservatorio Astrofisico di Torino, Via Osservatorio, 30, I-10025 Pino Torinese (Torino), Italy}

\maketitle
\section*{abstract}
Neutron energy spectrum for energies extending up to 15 MeV has been measured in the Hall A of INFN Laboratori Nazionali del Gran Sasso in Italy, by using a $\mathrm{1.5~m^3}$ stainless-steel tank filled with 1.2 ton Gd-doped liquid scintillator.
A two-pulse signature, due to proton recoils and delayed  gammas from neutron capture on Gd, has been used for event selection.
The neutron energy spectrum has been obtained from the detected proton-recoil spectrum, after unfolding the detector response, which had been simulated with GEANT4.
Although this spectrum has been measured previously \cite{12,13,cribier,haffke,best,aleksan,rindi}, the present work represents an improvement in the energy resolution with respect to previously published results.
The results obtained with this technique are compared with other measurements existing in literature.

\section{Introduction}
\label{intro}
The INFN Laboratori Nazionali del Gran Sasso (LNGS) is the largest facility in the world for underground astroparticle physics \cite{1}.
It is located in the center of Italy at an average depth of 3600 m w. e., where the cosmic ray flux is suppressed with respect to sea level, and is a suitable place for hosting experiments looking for rare processes, such as neutrinos from gravitational stellar collapse, neutrino-less double beta decay and dark matter interactions.
Nevertheless, radiation backgrounds are still present, primarily originating from the natural radioactivity of the surrounding rock.
The knowledge of background is therefore essential for a careful design of detectors and their shielding \cite{2}.
Neutron transport in matter is mainly computed by simulation packages and both the composition and the geometry of the shielding are calculated on the bases of the impinging neutron flux, known from measurements, and the attenuation factor required to reach certain physics goal.

Aim of the present work is the measurement of the neutron flux inside one of the experimental halls of the LNGS which could be used as input in simulation works for designing future large-scale detectors.
Exploiting the existence of a $\mathrm{1.5~m^3}$ Gd-loaded scintillator detector, partially exposed to the cavern rock radioactivity, a cost effective measurement of the neutron flux was possible.
The big volume of the Gd-doped scintillator detector allows the identification of neutron events by their capture on Gd, which is characterized by the emission of a $\mathrm{\gamma}$-cascade with enough energy ($\sim$8 MeV) to emerge from background.
During the thermalization process preceding its capture, the neutron hits protons of the scintillator and by studying the signals induced by proton recoils the original neutron spectrum can be reconstructed.

This paper describes the detector and its location inside the experimental hall A in section \ref{setup}.\\
Calibration procedures for both, electron and nuclear recoils, are described in section \ref{cal}.\\
Finally, event selection and unfolding method are discussed in section \ref{sec:4} and \ref{spect}, respectively.

\section{Experimental Setup}
\label{setup}
This work has been carried out in the framework of the LVD (Large Volume Detector) experiment, which is a neutrino observatory 
monitoring the galaxy since June 1992 to detect neutrinos from core collapse supernovae \cite{3}.
\begin{figure}[ht]
 \begin{center}
   \includegraphics[width=18pc,angle=0]{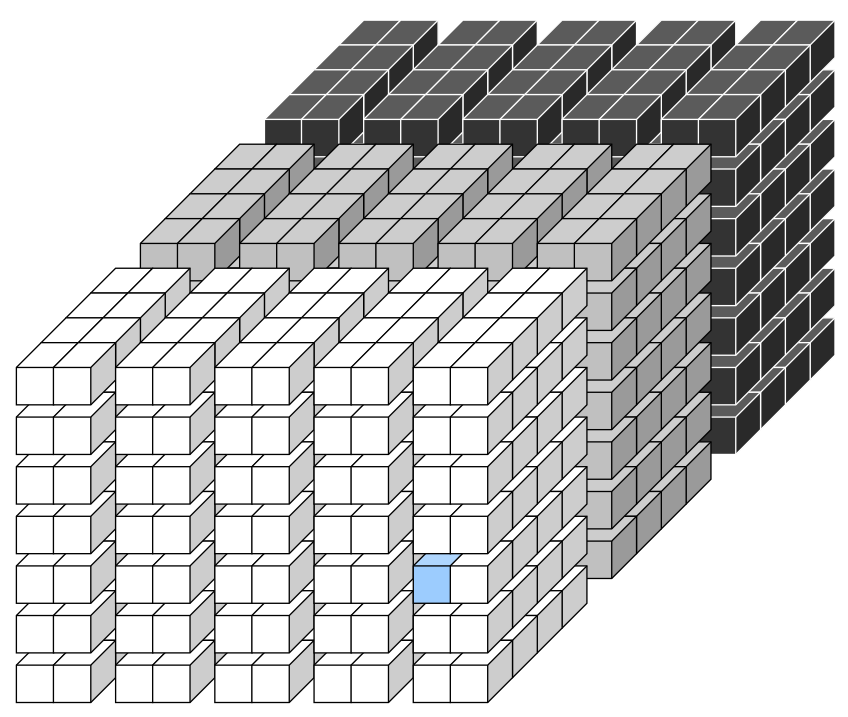}
   \caption{\label{cubetti} Drawing of the detector, represented in bluish, embedded in the LVD experiment, southern side.}
 \end{center}
\end{figure}
LVD is placed in the Hall A of the Gran Sasso Laboratory and
is made by an array of 840 scintillator detectors
$\mathrm{1.5~m^3}$ each. 
One of these, highlighted in figure \ref{cubetti}, has been adopted to perform the measurement reported in this paper.
It is placed on the southern region of the array and has several differences compared with the typical LVD detector, regarding scintillator, photomultipliers, front-end electronics and data acquisition.

The liquid scintillator of the standard LVD detector is a mixture of aliphatic and aromatic hydrocarbons ($\mathrm{C_nH_{2n}}$ with n = 9.6), also known as \emph{White Spirit} or \emph{Ragia Minerale}. 
The liquid scintillator of the counter we used for these measurements contains about 8\% of aromatics, and is doped with 1 g/l of PPO as activator and 0.03 g/l of POPOP which acts as wavelength shifter.
This scintillator has been developed several years ago by the component of the LVD Collaboration from the Institute for Nuclear Research of the Russian Academy of Sciences \cite{4}.
Years later, in the frame of a collaboration between the LVD and MetaLS (Metal and Rare Earth doped Liquid Scintillator) groups \cite{5}, this counter was doped with $\mathrm{[Gd]=0.93~g/l}$ (about 0.115\% in weight). 
Optical properties and characteristics of this scintillator can be found in \cite{5}. 
In particular the attenuation length at 425 nm, corresponding to the maximum sensitivity of the photocathode of the used photomultipliers, remains greater than 10 m after Gd doping, well above the detector dimensions $\mathrm{(1x1x1.5)~m^3}$.

Organic scintillators contain a relatively large number of hydrogen atoms per atom of carbon. They are therefore efficient moderators and are potentially suitable for fast neutron spectroscopy as well.
On the other hand,
doping the scintillator with Gd enhances the capability of tagging neutron captures, 
with respect to n-captures on protons,
resulting in a shorter average capture time and a higher energy emitted by the capture process, both contributing to significantly increase the signal to noise ratio.
At a Gd concentration of 0.93 g/l, the detector average thermal n-capture time is $\mathrm{\approx 27 \mu s}$ \cite{5}, about 7 times shorter than that of a standard LVD counter \cite{6}. At the same time the total energy emitted by n-capture on Gd is about 8 MeV, instead of the 2.2 MeV $\mathrm{\gamma}$ quantum for the $\mathrm{p(n,\gamma)D}$ reaction.

The detector has been equipped with 3 5" photomultipliers (PMTs) Photonis XP3550, that have a good gain stability over time. The PMTs have been tested and calibrated by measuring the single photoelectron spectrum (SPE) with a light emitting diode (LED) and their gain has been monitored for years.
Each PMT is connected to a 14 bit, 40 MHz bandwidth and 100 MS/s waveform digitizer (CAEN V1724).
The trigger condition is the three-fold coincidence among the PMTs. 
The trigger threshold has been set low enough to have a good efficiency for the detection of gammas from neutron captures and will be discussed in more details in section \ref{sec:4}. 
Finally, to reject muon-induced neutron  events originated by the passage of muons inside the LVD experiment, the logical OR of all triggers coming from the LVD counters surrounding the detector is also acquired.
\section {Detector Response}
\label{cal}
The detector response has been studied for both electron and nuclear recoils. 
The former by using a $\mathrm{^{60}Co}$ gamma source and the latter by using an $\mathrm{AmBe(\alpha,n)}$ neutron source to characterize the light output response to the different incident radiations.
\subsection{Electron recoils}
\begin{figure}[htpb]
 \begin{center}
  \includegraphics[width=21pc,angle=0]{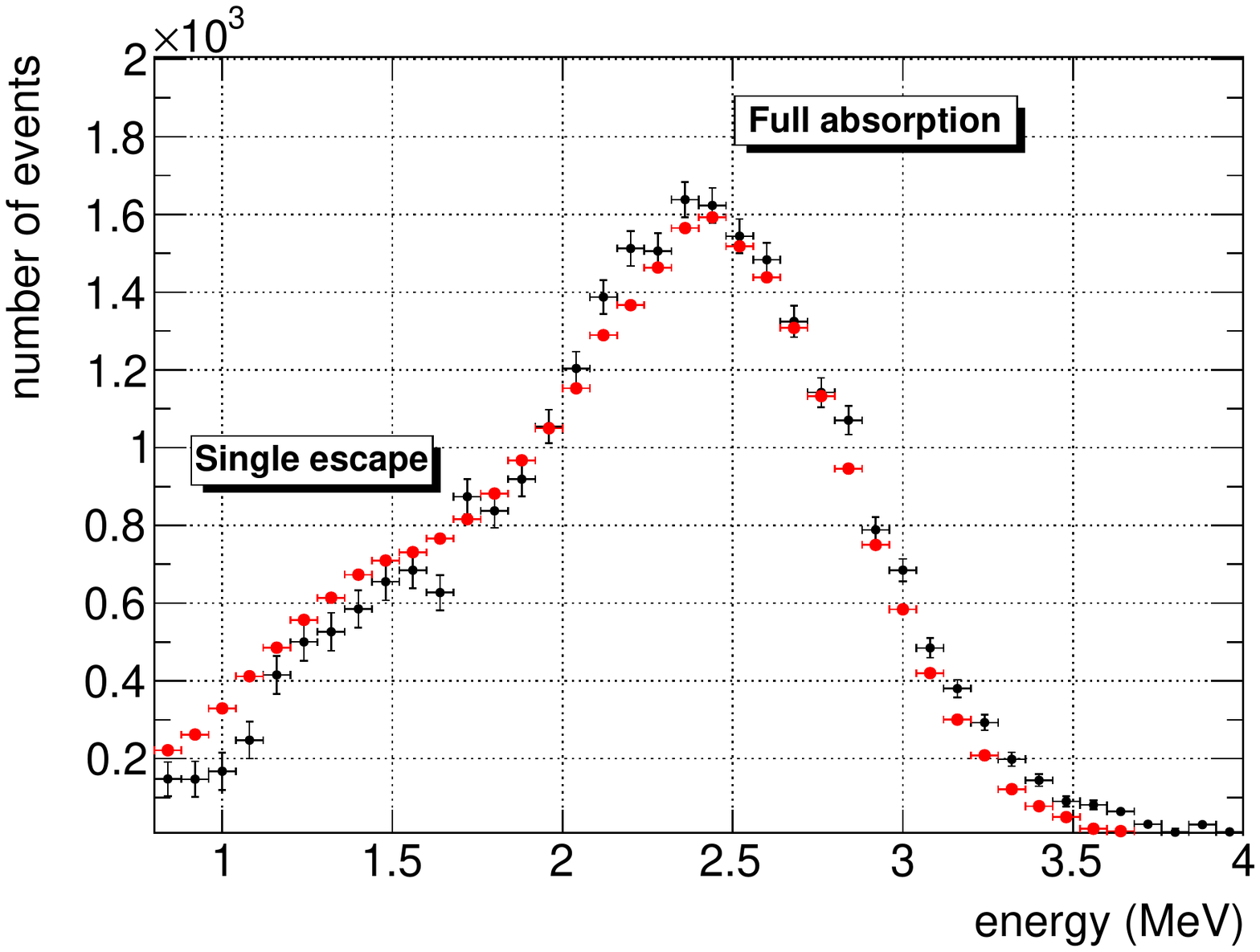}
  \caption{\label{cocalib} Typical spectrum from $\mathrm{^{60}Co}$ calibration: black and red dots represent experimental data and Monte Carlo simulation respectively. The energy scale has been established with 3$\%$ accuracy. In addition to the full absorption peak, corresponding to events completely contained in the active volume of the detector, we can distinguish a bump of events corresponding to one of the two gammas escaping the detector.}
 \end{center}
\end{figure}

Energy calibration of the detector has been performed at the beginning of the data taking (June 2011) and repeated at the middle (March 2012) and at the end (November 2012), using a $\mathrm{^{60}Co}$ source. The maximum gain difference is 3.4$\%$, no indication of degradation over time is observed.
The $\mathrm{^{60}Co}$ decays with a half life of $\mathrm{t_{1/2}=5.27~y}$ emitting two gammas of energy $\mathrm{E_1=1.17}$ and $\mathrm{E_2=1.33}$ MeV. 
The source has been inserted inside the detector using a stainless steel rod and was kept in the center of the scintillator volume during the calibration run.
This operation was made under a constant Argon stream to avoid any contact of the scintillator with the oxygen of the atmosphere.
The spectrum obtained after background subtraction, summing up the contribution of all PMTs, is shown in figure \ref{cocalib} with black dots, while a summary of the results is shown in table \ref{cosummary}. \\
The mean free path of 1.25 MeV photons in liquid scintillator is about 20 cm, which corresponds to a probability for a gamma-ray generated in the center of the detector to be completely contained in the active volume of about 95\%. 
The remaining 5\% of events contributes to the bump visible in figure \ref{cocalib} from 1 to 1.8 MeV.
\begin{figure}[ht]
 \begin{center}
   \includegraphics[width=21pc,angle=0]{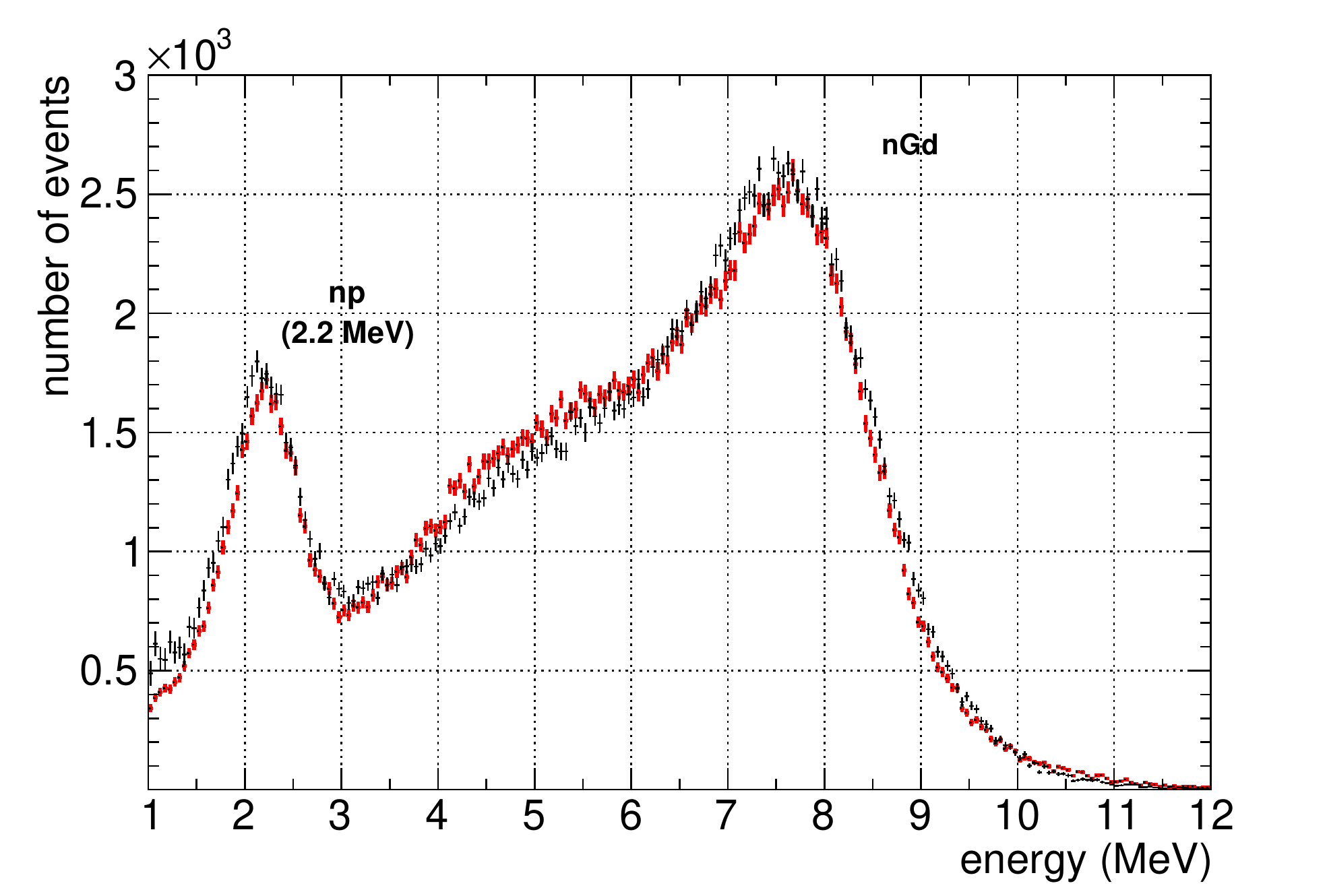}
\caption{\label{speconf2}Energy spectrum of signals due to neutron captures after background subtraction (in black), compared with MC simulation (in red).}
 \end{center}
\end{figure}
This calibration procedure allows us to establish the energy scale with a 3$\%$ uncertainty, dominated by systematics, ensuring that the detector gain remained constant, within uncertainty, during the whole period of data taking.
\begin{table}[htpb]
 \begin{center}
   \begin{tabular}{|l|l|}
 \cline{1-2}
\textbf{Date}    & \textbf{pC/MeV}  \\  \cline{1-2}
Jun. 2011 & 26.3$\pm$0.8    \\  \cline{1-2}
Mar. 2012 & 27.2$\pm$0.8    \\  \cline{1-2}
Nov. 2012 & 26.7$\pm$0.8    \\  \cline{1-2}
   \end{tabular}
  \caption{\label{cosummary} Summary of calibration factors measured during the data taking.}
 \end{center}
\end{table}
Moreover, the main crucial parameters, i.e., PMT high voltage, Radon activity, temperature and humidity of the environment were also continuously monitored.\\
The MC simulation of the detector response based on the Geant4 toolkit \cite{7} has been developed including the position of nearby LVD counters (details can be found in \cite{8}).
Data and MC are shown in figure \ref{cocalib}, with black and red dots, respectively.

\subsection{Nuclear recoils}
\label{ncalibsection}
To calibrate and monitor during data taking the detector response to nuclear recoils, an $\mathrm{AmBe(\alpha,n)}$ source was used, inserted inside the detector using a stainless steel rod and kept in the center of the scintillator volume as the $\mathrm{^{60}Co}$ one.

This source emits about 10 neutrons per second, and the emitted neutrons are thermalized by elastic scattering on the hydrogen atoms of the organic scintillator. 
During the slowing down process, a fraction of the neutron kinetic energy is converted to photons which can be detected by the PMTs giving a prompt signal. 
After being thermalized, neutrons are eventually captured by Gd and H (and with lower probability by C and Fe nuclei) producing a delayed signal. 
\begin{figure}[ht]
 \begin{center}
   \includegraphics[width=21pc,angle=0]{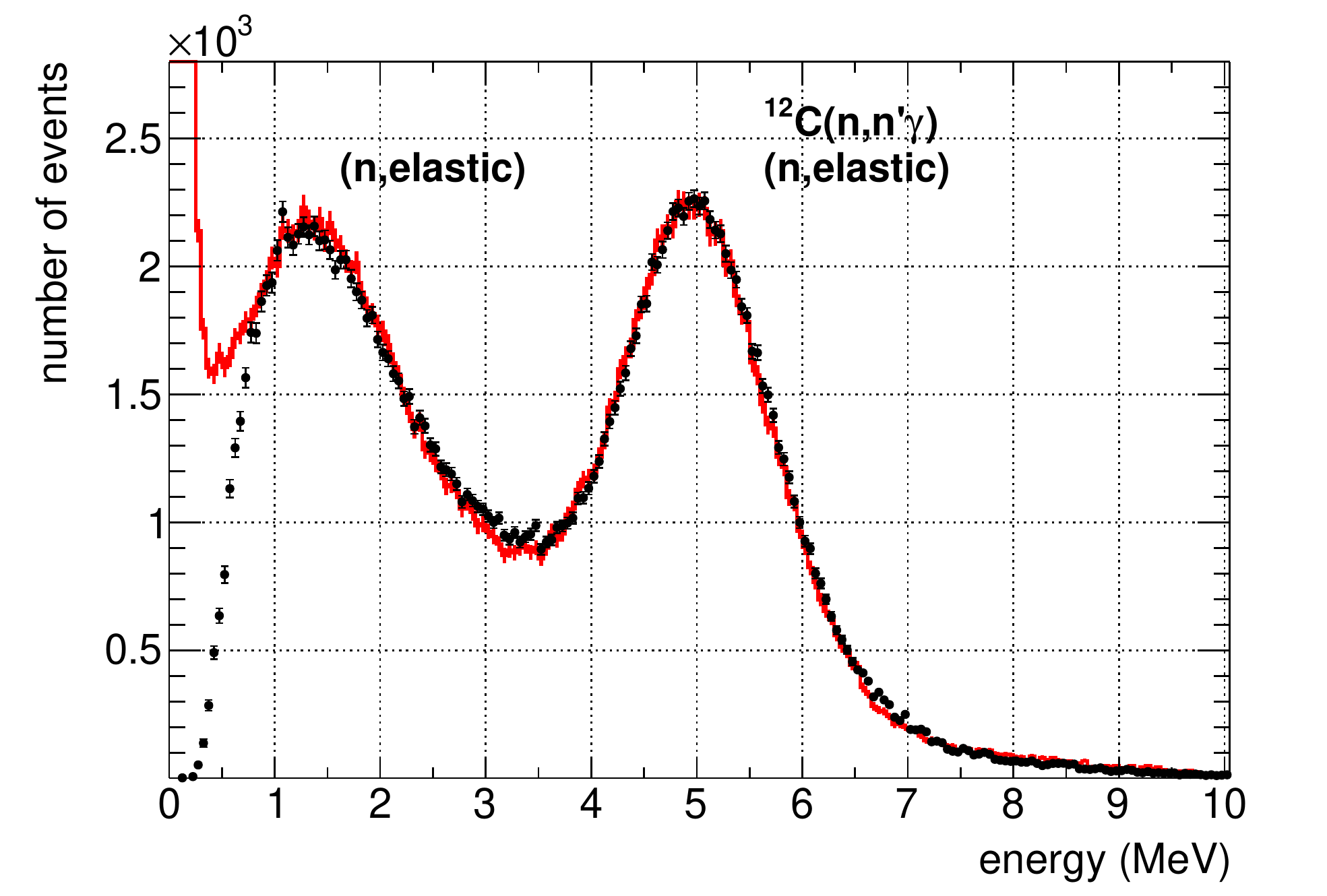}
\caption{\label{speconf} Energy spectrum of the prompt signals, after background subtraction (in black), compared with the simulation (in red). The detection threshold is at about 0.8 MeV.}
 \end{center}
\end{figure}
Neutron captures on H are followed by the emission of a 2.2 MeV $\mathrm{\gamma}$ quantum from the deuterium de-excitation, while captures on Gd create $\mathrm{\gamma}$ cascades with total energy $\mathrm{\approx 8 MeV}$.
Both processes are accurately described by the MC simulation, see fig \ref{speconf2}, supporting the elctron recoil calibration.

The mean n-capture time can be measured by studying the distribution of the time intervals between the prompt signal, due to the neutron elastic scattering with H, possibly accompanied by $\mathrm{^{12}C^*}$ de-excitation gamma-ray of energy 4.44 MeV, and the n-capture signal.
The mean n-capture time, strongly depends on the characteristics of the detector, mostly the Gd concentration (n-capture cross section), in our case 0.115\% in weight, and the detector geometry. 
The simulation predicts a mean n-capture time $t_{MC}=26.7\pm0.2$ $\mu s$, which is in agreement with the experimental value $t_{exp}=26.9\pm0.2$ $\mu s$, and also in agreement with previous measurements and simulations \cite{5}.
This quantity allowed us to monitor the scintillator stability during data taking in terms of Gd concentration (in solution) and light transmittance.

Finally, from the energy spectrum of the prompt signals, and the known energy spectrum of the neutrons emitted by the AmBe source \cite{9}, we were able to measure the light output response to neutron-induced nuclear recoils during the same years and with the same detector described in the present work. Details of this measurement can be found in \cite{8} where the quenching factor was found to be in good agreement with Birks semi-empirical model\cite{birks} with a Birks factor equal to 0.0140 $\pm$ 0.0007 g cm$^{-2}$ MeV$^{-1}$.

In figure \ref{speconf} is shown the energy spectrum of the detected prompt signals (after background subtraction) from the AmBe source (black dots), compared with the MC simulation (red dots). Note that the energy scale in abscissa is expressed in terms of MeV (electron equivalent), the same unit as in figure \ref{cocalib}.
The excellent agreement between the measured and simulated spectra validate the MC simulation and give us the confidence to use it for the reconstruction of the neutron flux from the cavern rock.

\section {Experimental Method}
\label{sec:4}
The key idea of this work is identifying neutron events by triggering on the neutron capture on Gd, which has enough energy to emerge from the background, then, by looking backward in time, to possibly find the proton recoil signal induced by the same neutron.
As a result, the neutron energy spectrum will be obtained by unfolding the proton recoil spectrum.

The three PMTs waveforms are continuously written in a circular memory buffer.
When a trigger occurs, High Threshold trigger (HT from now on), a time window of $\approx 640 \mu$s around the trigger signal ($\mathrm{2^{16}}$ samples for each PMT) is frozen. 
The DAQ software supports the Zero Length Encoding (ZLE) algorithm, providing data transfer in compressed mode.
The ZLE threshold, which we will refer to as Low Threshold (LT) since it is lower than the trigger threshold, can be set independently on each input channel.
The lower is LT the less is the energy of the neutrons we can detect. 
On the other hand HT has consequences on the detection efficiency but does not affect the energy range of the neutron spectrum.

HT has been set to have 100\% efficiency at 4.0 MeV, and LT to have 100\% efficiency at 1.0 MeV. 
In these conditions the trigger rate is approximately 0.8 Hz, therefore the detector runs with almost zero dead-time.

\begin{figure}[ht]
 \begin{center}
   \includegraphics[width=21pc,angle=0]{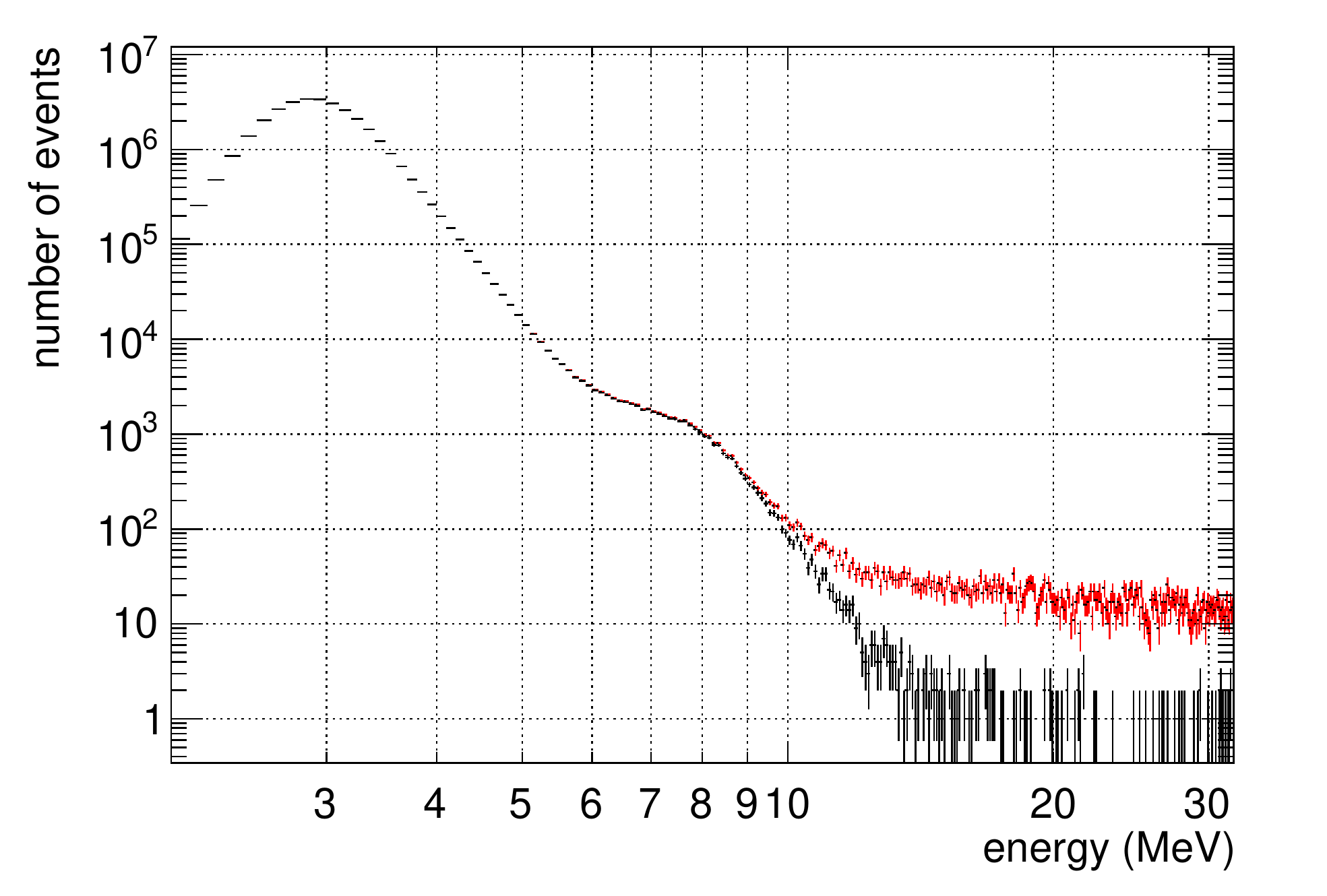}
   \caption{\label{3spec} Energy spectrum of all the trigger pulses (red line), pulses not tagged as muon (black line).}
 \end{center}
\end{figure}

In figure \ref{3spec} the energy spectrum of trigger signals is shown.
The red histogram reports the energy distribution of all HT trigger pulses, the black histogram has been obtained rejecting "muon" candidates from the red one, i.e., events in coincidence with at least one of the 280 LVD counters which belongs to the same \emph{LVD tower} \cite{3}.
The bump at about 8 MeV corresponds to the maximum of the gamma emission due to neutrons captured by Gd. 

It can also be that the recoil proton has enough energy to trigger the data acquisition by itself, in this case the signal due to the n-capture will be following the trigger.
To distinguish between the two signals of the pair the whole story of the event is needed. 
Therefore an acquisition window of $\mathrm{640 \mu s}$ is used with the trigger in the center, in order to have $\mathrm{320 \mu s}$ in both pre-trigger and post-trigger regions.\\ 
Since the mean n-capture time is about $\mathrm{27 \mu s}$, and considering a time delay of about $\mathrm{10 \mu s}$ for thermalizing, more than 98\% of neutrons are expected to be captured in the $\mathrm{120 \mu s}$ closest to the trigger, that is the region we use for the proton recoil search. 
The remaining part of the waveform is used to measure the background.

\begin{figure}[ht]
 \begin{center}
   \includegraphics[width=21pc,angle=0]{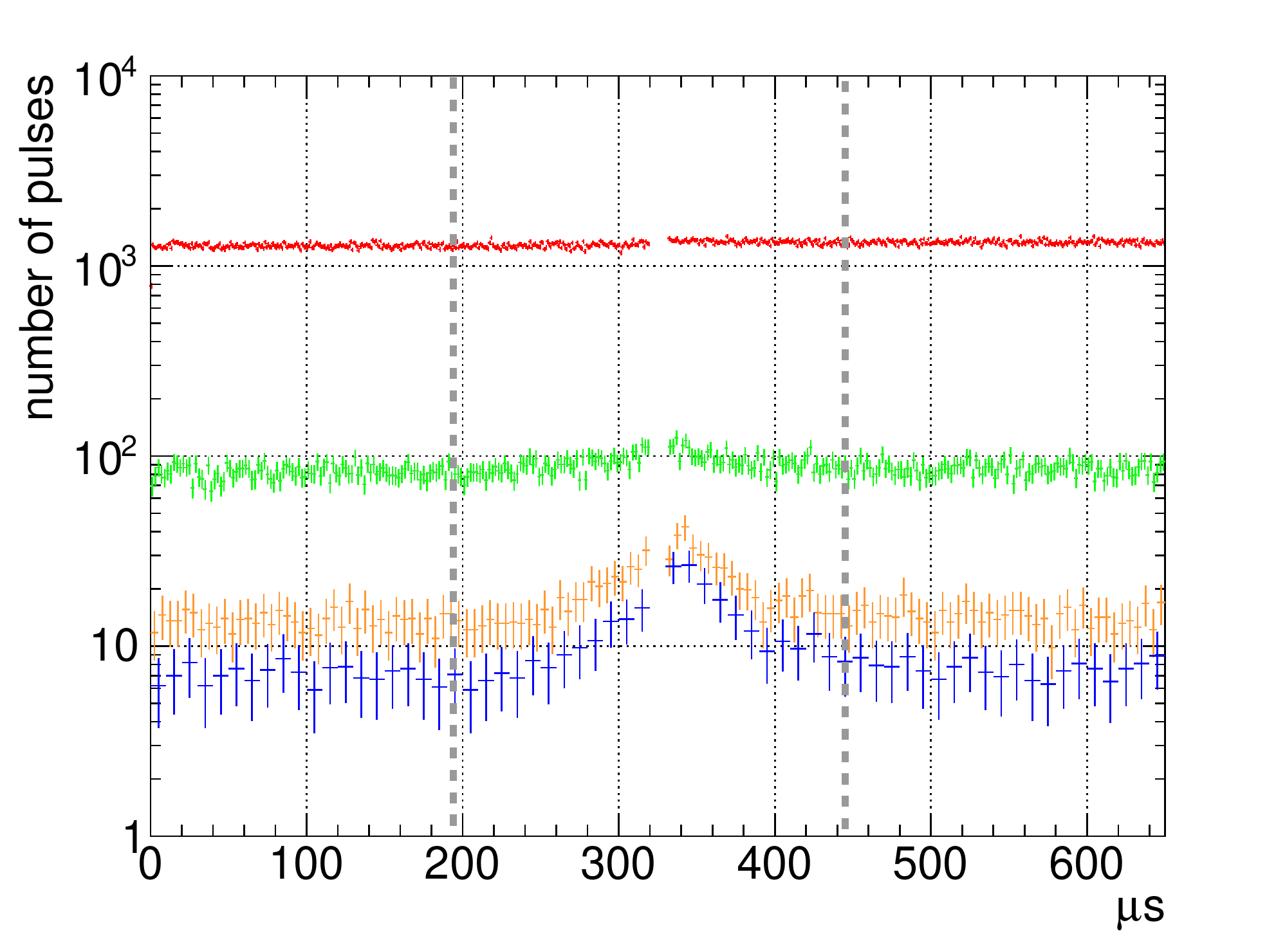}
   \caption{\label{dt} Arrival time of pulses in the neighbourhood ($\pm$320 $\mu s$) of the trigger signal. Different colours represent different trigger thresholds: 3 MeV (red), 4 MeV (green), 5 MeV (orange) and 6 MeV (blue). Dashed grey lines separate the signal region (in the center) from the background ones.}
 \end{center}
\end{figure}

In figure \ref{dt} the arrival time of pulses in the acquisition window is shown for four different trigger thresholds ranging from 3 to 6 MeV, from top to bottom in the figure. 
Triggers are not shown to focus the attention on the shape of the distributions, hence the missing points at $\mathrm{320 \mu s}$.
The distribution of the arrival time of the pulses is expected to have a flat component due to background signals (being uncorrelated to the trigger), while an exponential distribution is a distinguishing feature of a neutron event in which one pulse is due to the proton recoil and the other one to the n-capture. 
The exponential distribution emerges from the background when the trigger threshold is greater than 4 MeV.
By fitting the right and left side of the distribution, the exponential slopes have been found equal to (24 $\pm$ 3) $\mu$s and (28 $\pm$ 3) $\mu$s which are in good agreement with the value predicted by the MC (26.7 $\pm$ 0.2 $\mu$s), supporting that, at this energy threshold, the detector can discriminate between neutron events and gamma background.

\section{Neutron Energy Spectrum}
\label{spect}
In this work, we use data from June 2011 to October 2012: over this period the detector was active for about 485 days, corresponding to 97$\%$ livetime.
Neutron induced nuclear recoils are selected requiring (at least) 1 pulse of energy greater than 1.0 MeV (electron equivalent) occurring within $\mathrm{\pm}$ 120 $\mu s$ from the trigger signal, which has to be $\mathrm{\geq 5~MeV}$.
Whenever the time difference exceeds this limit, the pair of pulses is considered as uncorrelated and therefore due to background.

\begin{figure}[htpb]
 \begin{center}
  \includegraphics[width=20pc,angle=0]{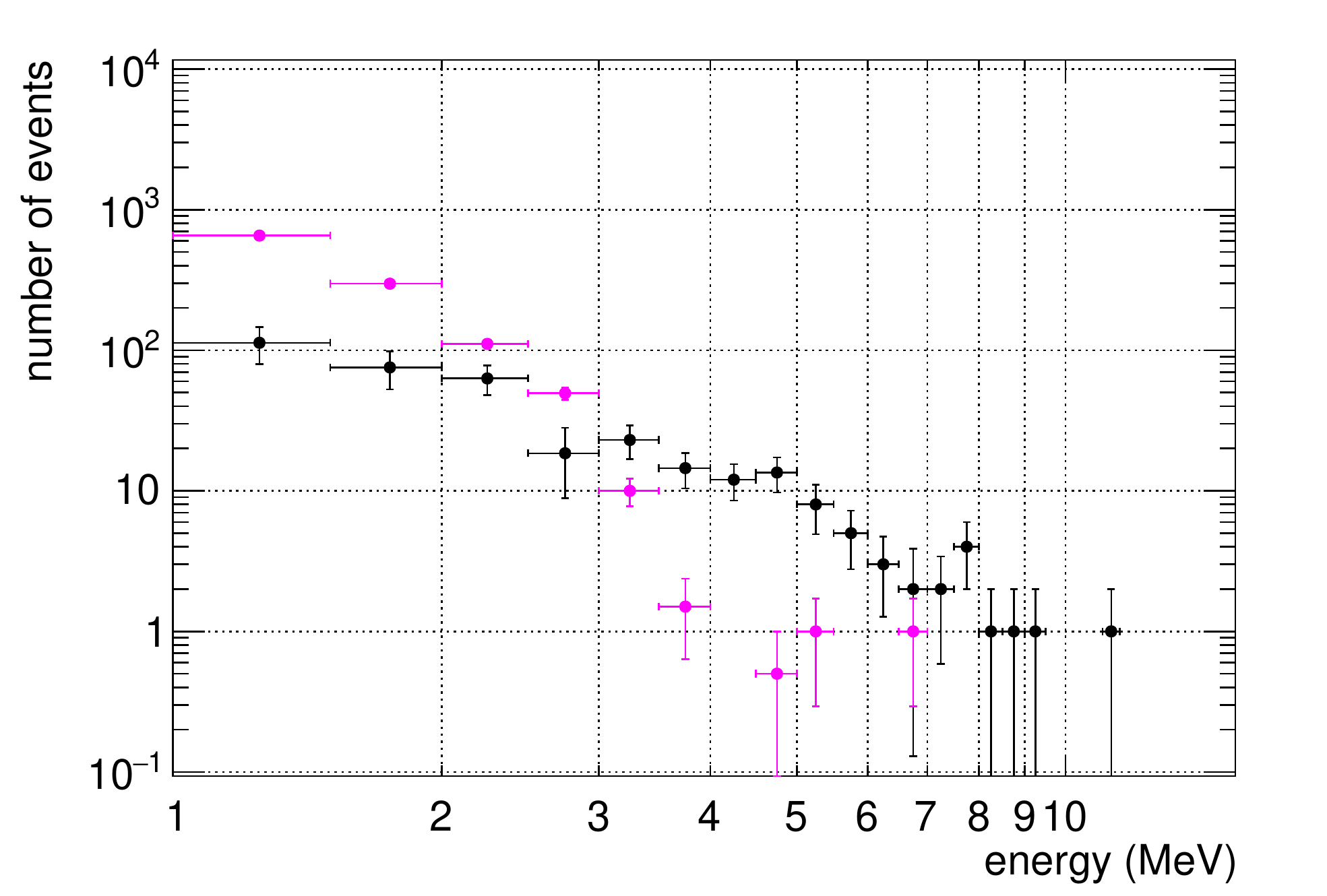}
  \caption{\label{erecoil} Energy spectrum of the recoil candidates after background subtraction (black dots), and background spectrum (magenta dots).}
 \end{center}
\end{figure}

Muon-induced neutron events occur with a probability depending on both the muon energy\footnote{The average energy of muons reaching the LNGS is about 270 GeV \cite{10}.} and the mass of the target nucleus. Therefore their multiplicity as well as their energy depends on the medium the muon is passing through. Ideally here we want our result to be experiment independent, therefore neutrons produced by the passage of muons inside LVD will be discarded from the analysis. Also we have to consider that the prompt signal of these events is partially due to the energy deposited by the muon passing through the scintillator and partially to the energy deposited by the neutron scattering off nuclei.\\
Since the neutron spectrum is reconstructed from prompt signals, to avoid a systematic overestimation of the neutron energy we decided to reject all the events in time coincidence with muons detected by LVD. Due to the low rate of the LVD detectors the dead time introduced is negligible.\\
In this way we reject most of the neutrons produced by muons inside the experiment.
Veto inefficiency can be derived from geometrical arguments and indeed an estimation of this number has been already published in \cite{effi}, where
the upper limit to the flux ($E_n$ $>$ 10 MeV) for neutrons produced inside LVD by undetected muons, (hence surviving to the LVD veto) is less than $3.6\cdot10^{-11}$ $cm^{-2} s^{-1}$.
This is more than one order of magnitude less than the neutron flux measured in this work.
During 485 live days, after muon rejection and background subtraction, we found 360 nuclear recoil events. 
The energy distribution of these events is shown in figure \ref{erecoil} in black, compared with the background in magenta.
The excess of events from 4 to 5 MeV is attributed to 4.44 MeV de-excitation $\mathrm{\gamma}$-rays from the inelastic scattering $\mathrm{^{12}C(n,n')}$ whose contribution is anyway taken into account by the MC simulation and included in the response functions of the detector.

The method used to evaluate the neutron flux $\mathrm{\Phi_n}$, from the nuclear recoil energy distribution, is based on the MC simulation of the detector response to mono-energetic neutrons. 
This simulation describes accurately both electron and nuclear recoil interactions in this same detector, see figure  \ref{cocalib}, \ref{speconf2}, \ref{speconf}, and \cite{8},  when radioactive sources are placed in its center.
Now the point source is replaced in the simulation by an isotropic neutron field, i.e., monochromatic neutrons are produced randomly on a spherical surface whose center coincides with the center of the southern face of the detector and the radius is equal to 4.5 m.
In this geometry 2$\pi$ of the sphere are entirely embedded in LVD and almost no neutrons from cavern rock radioactivity can reach the detector from this side.
Cosmogenic neutrons can do it but we tag them using LVD as a veto.
Neutrons originating from the remaining 2$\pi$ of the sphere have higher probability to be detected, contributing to the vast majority of the events.
This side of the detector is directly exposed to the cavern walls, hence neutrons can reach it without being slowed down.
However, for neutrons entering the detector from south, the GERDA watertank has been identified as the major contributor to neutron moderation.
Therefore to study the attenuation due to the GERDA watertank an additional simulation has been run using a spherical surface of radius equal to 22 m which is big enough to contain it entirely. Its contribution corresponds to a reduction of the geometrical efficiency, defined hereafter in text, by 30$\%$.

Let $\mathrm{N_{gen}}$ be the total number of neutrons generated by the simulation, $\mathrm{N_{imp}}$ the number of neutrons reaching the active volume, and $\mathrm{N_{det}}$ the number of neutrons which satisfy the selection criteria (giving at least 1 MeV electron equivalent prompt signal and at least 5 MeV from the following capture) intrinsic and absolute efficiencies can be defined as:

\begin{equation}
  \begin{aligned}
    \epsilon_{int}(E_n) & = \frac{N_{det}}{N_{imp}}
  \end{aligned}
    \label{int}
\end{equation}

\begin{equation}
  \begin{aligned}
    \epsilon_{abs}(E_n) & = \frac{N_{det}}{N_{gen}}
  \end{aligned}
  \label{abs}
\end{equation}
In particular the absolute efficiency can be expressed as: $\epsilon_{abs}(E_n) =  \epsilon_{int}(E_n) \cdot  N_{imp}/N_{gen}$  ,
where the last term is the geometrical efficiency, which is the ratio between the number of neutrons reaching the active volume of the detector, and the total number of neutrons generated by the simulation on the surface of the sphere of radius 4.5 m. It is independent on the energy of the neutron (at least in the energy range we are considering) and its value is: $(3.0 \pm 0.1) \cdot 10^{-3}$.
Multiplying by (1-0.3) to take into account the absorption due to the water-tank of the GERDA experiment, we obtain:  $N_{imp}/N_{gen} = \mathrm{(2.1 \pm 0.2) \cdot 10^{-3}}$.

\begin{figure}[htpb]
 \begin{center}
  \includegraphics[width=20pc,angle=0]{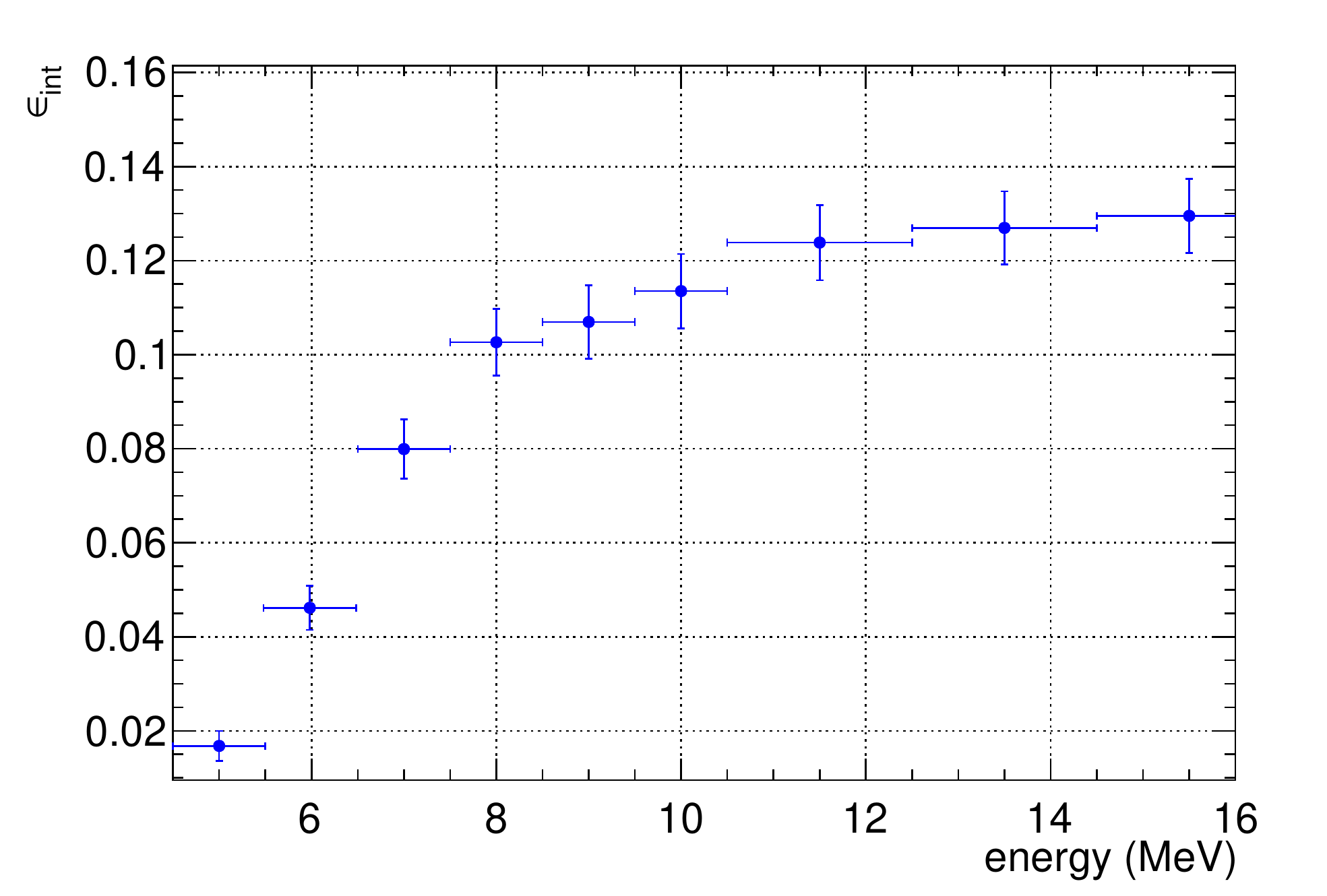}
  \caption{\label{effi} Intrinsic efficiency $\epsilon_{int}$ as a function of the neutron energy.}
 \end{center}
\end{figure}

The intrinsic efficiency (as well as the obsolute one) of our detector, strongly depends on the energy of the neutron and is shown in figure \ref{effi}.

Clearly, neutrons of energy equal to 5 MeV or less have a very low chance to give a prompt signal greater than 1 MeV electron equivalent, because of the scintillator light yield (quenching effect). This is the reason why, in the final result, the neutron spectrum starts at 5 MeV.

Considering the proton recoil spectrum of figure \ref{erecoil}, the number of counts $\mathrm{C_i}$ detected in the i-th energy bin is related to the incoming neutron spectrum $\mathrm{\Phi(E)}$ by a Fredholm integral equation of the first kind:
\begin{equation}
C_i = \int_{0}^{\infty} S_i(E)\Phi(E) dE
\end{equation}
where $\mathrm{S_i(E)}$ is the probability that a neutron of energy $\mathrm{E}$ deposits an energy between $\mathrm{E'}$ and $\mathrm{E'+\Delta E'}$ in the detector, corresponding to the i-th energy bin.
This equation can be written in discrete form:
\begin{equation}
\label{eqnflux}
C_i \approx \sum_n \mathcal{S}_{in}\Phi_n
\end{equation}
where $\Phi_n$ is the neutron flux in the n-th energy interval and $\mathcal{S}_{in}$ is a squared matrix whose rows are the 
visible-energy distribution for mono-energetic neutrons, i.e., the response functions.\\
The vector $\vec{\Phi}$ can be (uniquely) determined by minimization only if its length is equal to the rank of the matrix $\mathcal{S}_{in}$.
Since this is usually not the case for experimental data, which are affected by uncertainties, equation \ref{eqnflux} can not be solved exactly.
To avoid negative solutions and easily handle errors, the matrix inversion is performed using the least-squares approach, finding the vector $\vec{\Phi}$ that minimizes  $\left\lVert \mathcal{S}\vec{\Phi}-\vec{C} \right\rVert$.
\begin{figure*}
\begin{center}
\includegraphics[width=\textwidth]{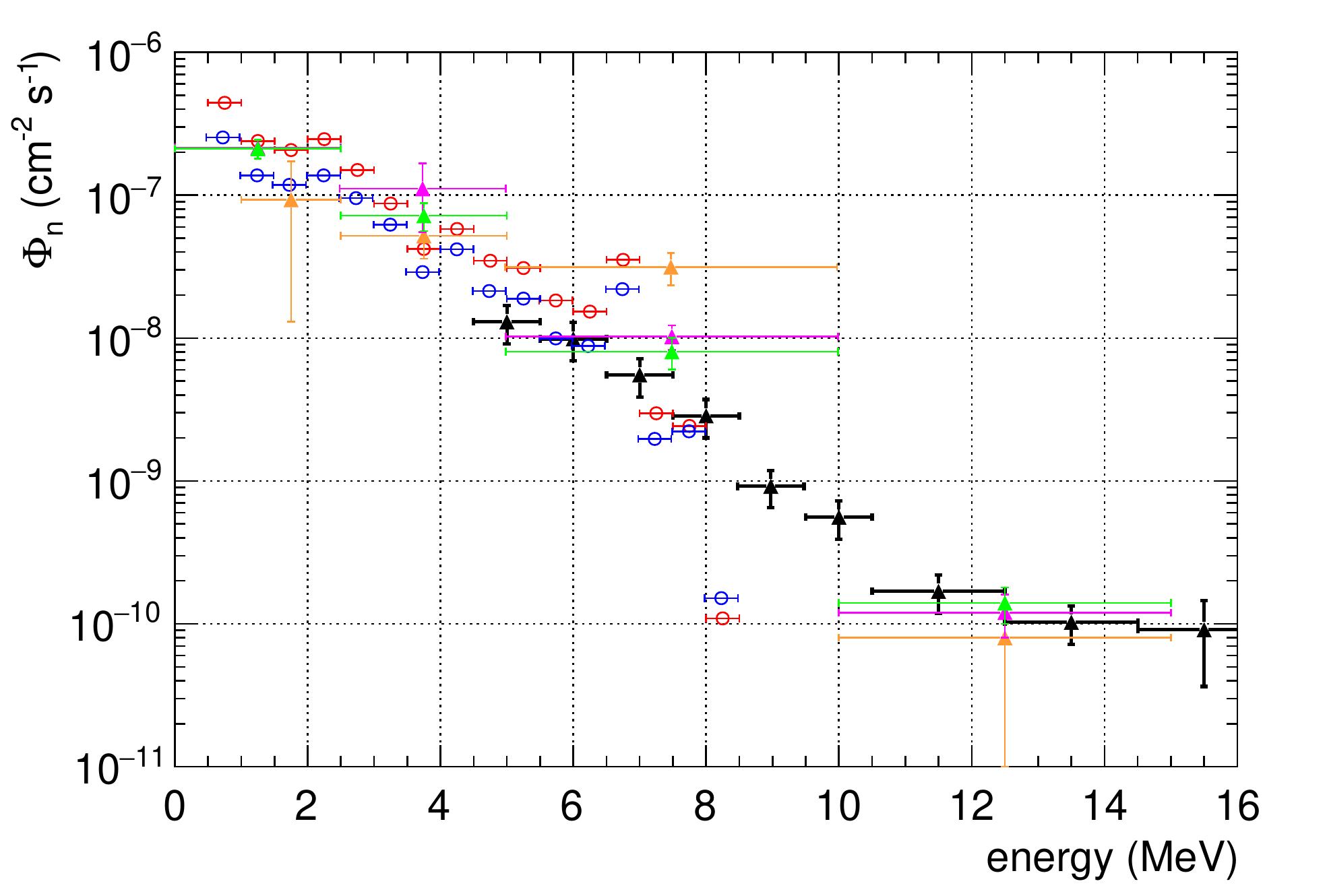}
\caption{\label{fig:8} Reconstructed neutron energy spectrum in the energy range from 5 MeV to 15 MeV (black triangles this work) compared with previous measurements represented by solid triangles: green and pink from \cite{13} and orange from \cite{12}.
Empty circles, in red and blue, represent the results of the simulations from \cite{11} of Hall A dry and wet concrete respectively.
Different energy bins are used by different authors, hence to make the comparison easier, each point in the figure shows the integral flux divided by the bin width.}
\end{center}
\end{figure*}

The result of the deconvolution (or unfolding) process is the spectrum $\vec{\Phi}$ shown in black in figure \ref{fig:8}, with purely statistical error bars, compared with other measurements and simulations existing in literature.\\
Systematic uncertainties, mostly due to simulation process are subdominant.
To double-check that the matrix inversion algorithm worked correctly, and to validate the considered error budget, we folded again the neutron spectrum with the response functions finding good agreement with the experimental data $\mathrm{C_i}$.

For energies $E_n$ $>$ 10 MeV the muon-induced neutron component becomes dominant, its integral flux obtained in this work is $\sim7.5\cdot10^{-10}$ $cm^{-2} s^{-1}$ which is in good agreement with the value quoted in table V of \cite{mei}, equals to $7.3\cdot10^{-10}$ $cm^{-2} s^{-1}$.

This component was not included by Wulandari, et al. \cite{11}, who simulated the flux of neutrons induced by radioactivity only, that's why their spectra (red and blue empty circles in figure \ref{fig:8}) end at about 8 MeV.
Measurements of the neutron flux, integrated over a broad range of energies, have been published along the years and can be found in: \cite{cribier}, \cite{haffke}, \cite{best}, \cite{aleksan} and \cite{rindi}. They are not shown in figure \ref{fig:8}.
\section{Conclusions}
\label{sum}
A 1.5 cubic meter detector located in the INFN Laboratori Nazionali del Gran Sasso (LNGS) filled with Gd-doped liquid scintillator has been used to measure the energy spectrum of neutron from rock radioactivity. 
The dimensions of the detector, allows the confinement of the entire event, and the strong signature of the neutron events, provided by n-capture on Gadolinium, allows a direct measurement of the energy spectrum of scattered protons.
Previous studies of the quenching effect, made on this same detector \cite{8}, allow a correct interpretation of this energy spectrum.
Deconvolution of kinematics and detection effects leads to a final reconstructed neutron spectrum obtained in the absence of any hypothesis on its shape.

The obtained results are limited to the neutron energy range between 5 and 15 MeV and take already into into account the shielding effect due to other experiments that occupy the experimental Hall.\\
At energies lower than about 5 MeV, mainly because of quenching, neutrons have a very low chance to give a prompt signal greater than 1 MeV electron equivalent that is our energy detection threshold for recoils.\\
For energies greater than about 10 MeV, on the other side, the contamination due to untagged muon induced neutrons becomes dominant. 
In this energy region our results can not be compared with MC simulations from \cite{11}, because this component was not considered by the authors being strongly dependent on how the experimental hall itself is filled.

\section*{Acknowledgments}
The authors wish to thank the LNGS Special Techniques Service for their support and cooperation. This measurement would not have been possible without the collaboration of our LVD colleagues, which hosted the neutron detector inside the LVD array which performed as a muon veto. Finally we want to thank the MetaLS group that realized the Gd doping of the scintillator.

\end{document}